\documentclass[a4paper,12pt]{article}
\usepackage[utf8x]{inputenc}
\usepackage{graphicx}
\usepackage{amssymb,amsmath}
\usepackage{textcomp}

\usepackage{ctable}
\usepackage{graphicx}
\usepackage{subfigure}
\usepackage{placeins}
\usepackage[english]{babel}
\usepackage{booktabs}
\usepackage{multicol}
\usepackage{caption}
\usepackage{cite}

\title{Properties of hadronic systems according to the non-extensive self-consistent thermodynamics}
\author{A. Deppman}
\date{Instituto de Física, Universidade de São Paulo - IFUSP, Rua do Matão, Travessa R 187, 05508-900 São Paulo-SP, Brazil}

\begin{document}

\maketitle

\begin{abstract}
The non-extensive self-consistent theory describing the thermodynamics of hadronic systems at high temperatures is used to derive some thermodynamical quantities, as pressure, entropy, speed of sound and trace-anomaly. The calculations are free of fitting parameters, and the results are compared to lattice QCD calculations, showing a good agreement between theory and data up to temperatures around 175 MeV. Above this temperature the effects of a singularity in the partition function at $T_o=$ 192 MeV results in a divergent behaviour in respect with the lattice calculation.
\end{abstract}

One of the challenges of nuclear physics nowadays is to understand the nature of the transition from confined to deconfined regimes of strong-interacting systems. In this regard the thermodynamics of hadronic medium has received much attention due to the possibility of describing the equation-of-state at temperatures just below the critical value, $T_c$, where the system could still be represented in terms of hadronic degrees of freedom. Quantities as the trace-anomaly and the speed of sound in hadronic medium are considered sensitive to the phase transition.

Hadron resonance gas (HRG) models~\cite{Broniowski1, Broniowski2,Chatterjee, Megias1, Arriola} use Hagedorn's theory~\cite{Hagedorn} to introduce an exponentially increasing mass spectrum for hadrons. This behaviour of the mass spectrum was considered to be necessary~\cite{Noronha} to describe the lattice-QCD results. However, a few problems have appeared: Hagedorn's theory fails to explain the experimental results from hadron-hadron collisions for energies above 10~GeV~\cite{Bediaga,Sena2}; different HRG models diverge on the value for the Hagedorn's temperature when fitting the available information for mass spectrum, some of them being well-above the critical temperature expected from lattice-QCD calculations~\cite{CWmass}.

Recently the non-extensive self-consistent thermodynamical theory was developed predicting a limiting effective temperature, $\tau_o$, a limiting entropic index, $q_o$, and a power-law mass-spectrum~\cite{Deppman}. This theory imposes a much more restrictive test to the role played by Tsallis statistics in High Energy Physics (HEP). Similar predictions about a limiting effective temperature were also proposed through a massless parton gas model~\cite{Biro_Peshier}. 

Despite being still a rather controversial subject, the use of non-extensivity in HEP has received increasing interest over the last years (see for instance~\cite{Bediaga, Sena2,Deppman,Biro_Peshier,Cleyman_Worku,Lucas,Wilk2012,1Alice1,5Alice2,8Alice3,4Atlas1,9cms1,11cms2,12cms3,3lhcb1,6lhcb2,10lhcb3} and references therein), and it has been shown that experimental data from high energy collisions indeed supports the conclusions from Refs.~\cite{Deppman, Biro_Peshier}. In this work, another important test of non-extensivity is performed. Using the informations of experimental data analysis, the non-extensive self-consistent thermodynamics is completely described. Then it is used to calculate thermodynamical functions which can also be calculated by lattice QCD methods. Although power-law mass spectrum was discarded within extensive (Boltzmann) thermodynamics~\cite{Noronha}, in this work the non-extensive self-consistent predictions are investigated and compared to lattice-QCD calculations. It turns out that 
there is a good agreement between thermodynamics and lattice calculations up to temperatures around 175~MeV.

The non-extensive self-consistent theory has two main ingredients~\cite{Deppman}: the non-extensive statistics and ii) the fireball self-consistency principle. This theory predicts the limiting temperature and the limiting entropic index which characterize strongly interacting systems. More relevant to the present work are the mass spectrum, $\rho(m)$, and the partition function, $Z(V,\beta)$ derived in that theory, which are given by~\cite{Deppman}
\begin{equation}
\rho(m)=\gamma m^{-5/2}e_q(\beta_o m)
\label{mass_spectrum}
\end{equation}
and
\begin{equation}
\ln[Z(V,\beta)]=\frac{\gamma V}{2\pi^2 \beta^{3/2}} \int_M^{\infty} m^{-1} \rho(m) e_q(-\beta m) dm + Z_1(V,\beta)\,,
\label{partition_function}
\end{equation}
where $e_q(x)$ is the q-exponencial function 
\begin{equation}
 e_q(x)=[1+(q-1)x]^{1/(q-1)}\,,
\end{equation}
$\beta=1/\tau$, with $\tau$ being the effective temperature, $\beta_o=1/\tau_o$, $M$ is the lowest mass for which the mass spectrum formula is compatible with the experimental information on hadronic states (see text below), and $\gamma$ is a constant. The second term on the right of Eq.~\ref{partition_function} is 
\begin{equation}
 Z_1(V,\beta)=\int_0^M m^{3/2}\rho'(m)e_q(-\beta m) dm\,,
\end{equation}
with $\rho'(m)$ being the mass spectrum valid below the mass $M$.

Different analysis of transverse momentum distributions from high-energy particle collisions have shown that the prediction of the limiting temperature and of the limiting entropic index are in accordance with experimental data~\cite{Bediaga, Sena2,  Sena1, Cleyman_Worku,  Lucas}, resulting in $\tau_o=$(60.7$\pm$0.5)~MeV, $q_o=$1.138$\pm$0.006 and $\gamma=(5\pm3)\times 10^{-3}$~Gev$^{3/2}$. Also, it was shown that $\tau_o$ and $q_o$  are related through the equation
\begin{equation}
 \tau_o=T_o+(q-1)\,c\,,
\label{Tq}
\end{equation}
where $c$ is a constant depending on the energy transfer between the source and its surroundings~\cite{Wilk2012}. Since $\tau \rightarrow T_o$ as $q \rightarrow 1$, $T_o$ is considered to be the Hagedorn's temperature. From a systematic analysis of experimental data it was found that $T_o=T_H=$(192$\pm$15)~MeV and $c=$-(950$\pm$10)~MeV~\cite{Sena1}. Observe that the Hagedorn's temperature obtained in this way is in agreement with the lattice-QCD results~\cite{Ejiri,Miura}.

The mass spectrum given by Eq.~\ref{mass_spectrum} has been compared with the observed data. It was shown that it gives an excelent description of the available data for masses going from the pion mass, $m_{\pi}$, up to $m=$ 2.5~GeV with values for $\tau_o$ and $q_o$ similar to those obtained in the analysis of transverse momentum distributions~\cite{Lucas}. This result allows one to identify $M=m_{\pi}$, therefore it is reasonable to conclude that $\rho'(m)\equiv 0$, and consequently $Z_1(V,\beta)=0$. Hence the non-extensive self-consistent theory can completely describe the experimental information about the hadronic system formed in high-energy collisions and about the observed hadron-mass spectrum.

In the following Greek letters are used to designate  thermodynamical quantities which are calculated according to the non-extensive self-consistent theory and Latin letters to designate those quantities calculated by lattice-QCD. From equations~\ref{mass_spectrum} and~\ref{partition_function} it is possible to calculate all thermodynamical functions. The fact that $[(q_o-1)(\beta-\beta_o)M]^{-1/(q-1)} \gg 1$ in the range of temperatures of interest is used to simplify the expressions for the thermodynamical functions presented below. It is useful to write Eq.~\ref{partition_function} in terms of the  Gauss' hypergeometric function, $_2F_1$, resulting (using $\zeta$ to indicate partition function)~\cite{Deppman}
\begin{equation}
 \frac{1}{V}\ln[\zeta(V,\beta)]=\beta^{-3/2}\,_2{\cal F}_1(q,\beta)\,,
\label{LZ2F1}
\end{equation}
where
\begin{equation}
 _2{\cal F}_1(q,\beta)=\frac{\gamma (q-1)}{2 \pi^2} \left(\frac{1}{(\beta-\beta_o) M (q-1)}\right)^{\frac{1}{q-1}}\, _2F_1\left[\frac{1}{q-1},\frac{1}{q-1};\frac{q}{q-1};\frac{-1}{(q-1) (\beta-\beta_o)M }\right]\,.
\end{equation}

From the equation above it is possible to get the pressure, $\Pi$, through the relation
\begin{equation}
\Pi=\frac{\tau}{V}\ln[\zeta(V,\beta)]\,
\end{equation}
resulting in
\begin{equation}
\begin{split}
 \Pi= \beta^{-5/2}\,_2{\cal F}_1(q,\beta)\,.
\label{nepressure}
\end{split}
\end{equation}

Using the above expression for $\Pi$ and the relation
\begin{equation}
 \sigma= \frac{\partial \Pi}{\partial \tau}
\end{equation}
the entropy density $\sigma$ can be obtained, resulting
\begin{align}
 \sigma=\frac{5}{2} \beta^{-3/2}\,_2{\cal F}_1(q,\beta)\,.
\label{neentropy}
\end{align}

The energy density, $\varepsilon$, can be calculated from Eq.~\ref{LZ2F1} by
\begin{equation}
 \varepsilon=\frac{\tau^2}{V}\frac{\partial\,}{\partial \tau}\ln[\zeta(V,\beta)]\,,
\end{equation}
resulting in
\begin{align}
 &\varepsilon=\frac{3}{2} \beta^{-5/2}\,_2{\cal F}_1(q,\beta)\,.
\label{neenergy}
\end{align}


\begin{figure}[!h]
       \centering
                 \subfigure[]{\includegraphics[width=0.47\textwidth]{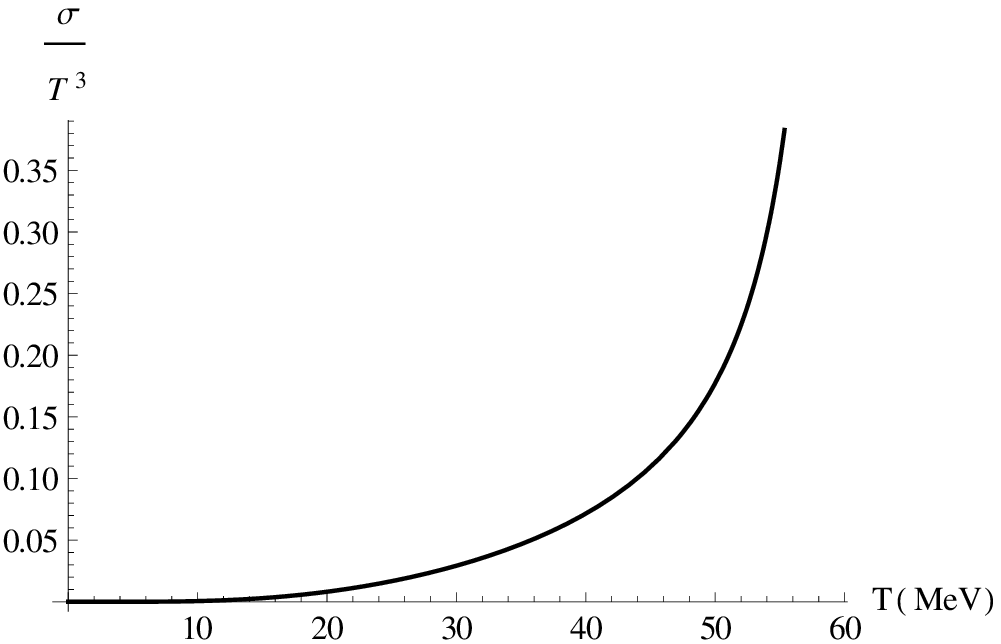}}
                \subfigure[]{\includegraphics[width=0.47\textwidth]{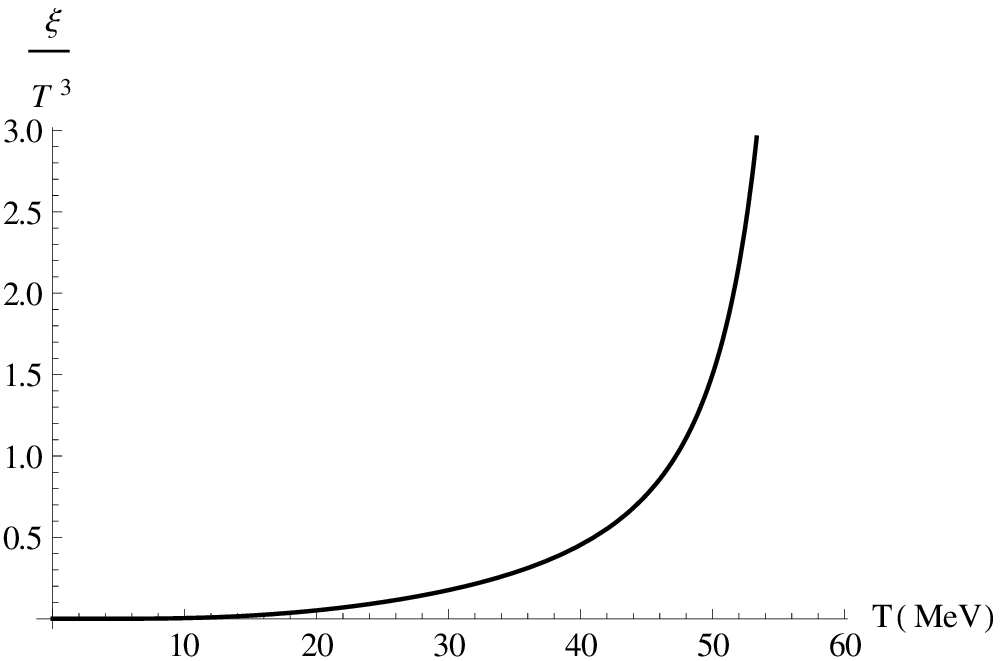}}
                       \caption{Non-extensive functions: normalized entropy density (a) and specific heat (b) as a function of the effective temperature. }
               \quad
       \label{fignese}
   \end{figure}

Other interesting quantities to be calculated are the so-called trace-anomaly, $\alpha(T)$, which is considered as an indication of phase transition, and the specific heat at constant volume. The trace-anomaly is calculated by
\begin{equation}
  \alpha(T)=\frac{\varepsilon-3\Pi}{\tau^2}=\tau\frac{\partial}{\partial \tau}\bigg(\frac{\Pi}{\tau^4}\bigg)\,,
\end{equation}
and the specific heat at constant volume by
\begin{equation}
 \xi=\frac{\partial \varepsilon}{\partial \tau}\,.
\end{equation}
Also, from the entropy density and specific heat it is possible to calculate the speed of sound in the hadronic medium, $\nu$, by
\begin{equation}
 \nu^2=\frac{\sigma}{\xi}\,.
\end{equation}

The results for the normalized entropy and specific heat are plotted in Fig.~\ref{fignese}. It is possible to observe that both quantities present a fast increase as the temperature approaches the critical value, $\tau_o$, which is related to the existence of a singularity in the partition function~\cite{Deppman}. This singularity is the cause of the limiting effective temperature in the thermodynamical non-extensive self-consistent theory.

\begin{figure}[!h]
       \centering
                 \includegraphics[width=0.47\textwidth]{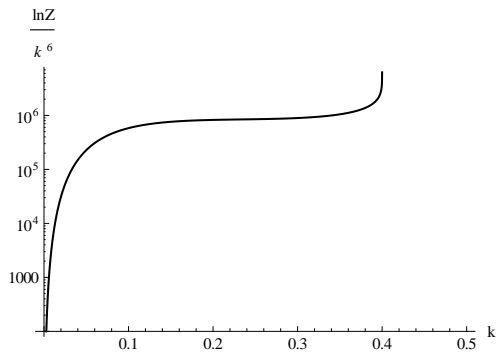}
           \caption{The function $k^{-6}\ln Z$ as a function of k. The flat region between $k=$0.1 and $k=0.4$ shows that $\ln Z$ is proportional to $k^6$. }
       \label{figk6lnz}
   \end{figure}

In order to compare the non-extensive calculations with the lattice calculations it is necessary first to find a map between the effective temperature, $\tau$, and the lattice-QCD temperature, $T$. Equation~\ref{Tq} establishes a connection between the critical temperature, $T_o$, and the critical effective temperature, $\tau_o$. This relation, however, cannot be used for arbitrary temperatures.
The map between $T$ and $\tau$ is expected to be a function $\tau(T)$ with the following properties:
\begin{enumerate}
 \item $\tau(T_o)=\tau_o$;
 \item $\tau(0)=0$;
 \item $\tau(T_1+T_2)=\tau(T_1)+\tau(T_2)$.
\end{enumerate}
The third property will be discussed below.

The function satisfying all those conditions simultaneously is
\begin{equation}
 \tau(T)=kT \equiv \frac{\tau_o}{T_o}T\,,
\label{Ttau}
\end{equation}
where the constant $k=\tau_o/T_o$ has been defined.

\begin{figure}[!h]
       \centering
                 \subfigure[]{\includegraphics[width=0.47\textwidth]{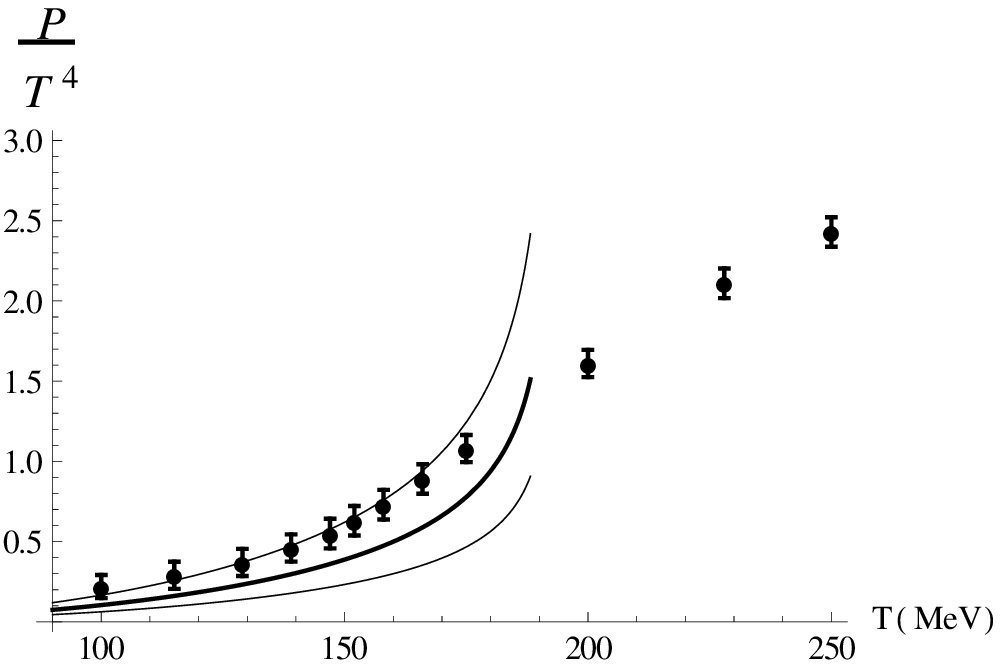}}
                \subfigure[]{\includegraphics[width=0.47\textwidth]{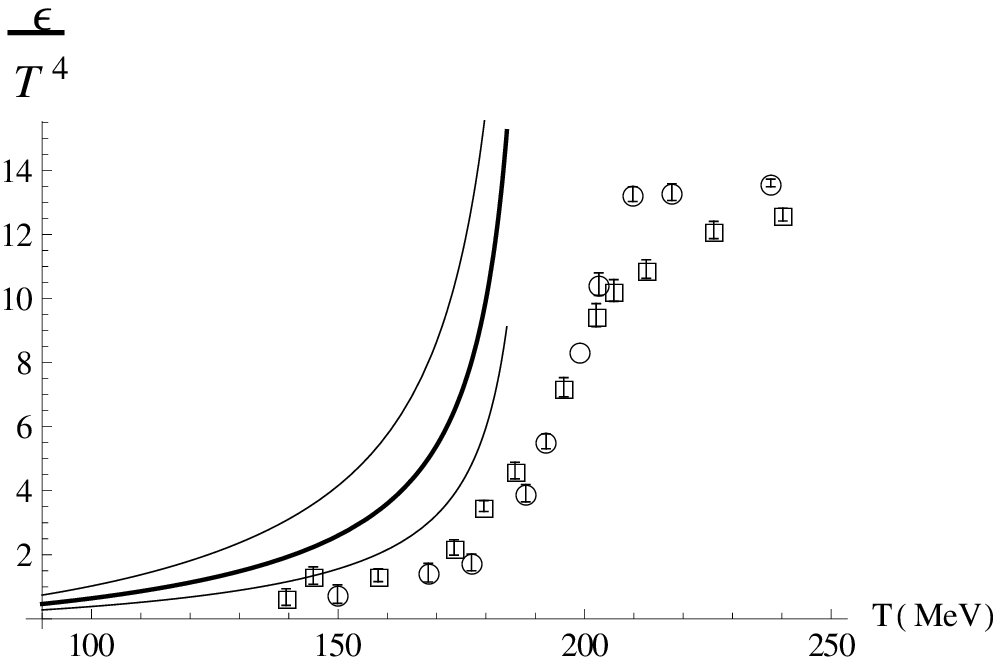}}
                       \caption{Normalized pressure  (a) and normalized energy density (b) as a function of the temperature. Thick line represents the theoretical results. Thin lines represent the upper and lower limits of the theoretical calculations considering the uncertainties in the parameter $\gamma$ of the mass spectrum. Full circles represent the results of lattice QCD from Ref.~\cite{Zoltan_Fodor}. Empty symbols are the results of lattice calculation from Ref.~\cite{F_Karsch}, with circles indicating p4, N$_t$=4, squares indicating p4, N$_t$=6.}
               \quad
       \label{figpa}
   \end{figure}

Note that Eq.~\ref{Ttau} determines also relations between other thermodynamical functions. In fact the partition function is transformed in such a way that
\begin{equation}
 \ln Z = k^{-6} \ln \zeta
\end{equation}
for $k$ around the value $\tau_o/T_o$, as can be observed in Fig.~\ref{figk6lnz}.

 With these relations established, it is now possible to compare the results from the non-extensive self-consistent theory with lattice calculations.

In the present work three sets of lattice data are used for comparison with the theoretical results, namely, that of S. Borsanyi et al.~\cite{Zoltan_Fodor} (L1), and those from M. Cheng et al~\cite{F_Karsch}(L2). The former is a 2+1 staggered quarks calculation with physical quark masses, stout action and $N_{t}$ in the continuum limit, while the calculations in the latter set are obtained with 2+1 staggered quarks with {\it almost} physical masses and p4 action for $N_{t}=$4 and $N_{t}=$6 . These three sets are not intended to provide a complete description of lattice-QCD results, but are representative of the state-of-art in the field.

\long\def\symbolfootnote[#1]#2{\begingroup%
\def\thefootnote{\fnsymbol{footnote}}\footnote[#1]{#2}\endgroup} 

The results for the normalized pressure are shown in Fig.~\ref{figpa}(a), and are compared to L1 results. Considering the uncertainties in the parameter $\gamma$\symbolfootnote[2]{One must keep in mind that the parameters in the non-extensive self-consistent theory are not free of correlations, therefore the uncertainty range presented here may be somewhat overestimated.} it is possible to observe a fair agreement between theory and lattice calculations up to $T\sim$185~MeV, the deviation thereof being due to the singularity of the partition function at the critical temperature.

\begin{figure}[!h]
       \centering
                 \subfigure[]{\includegraphics[width=0.47\textwidth]{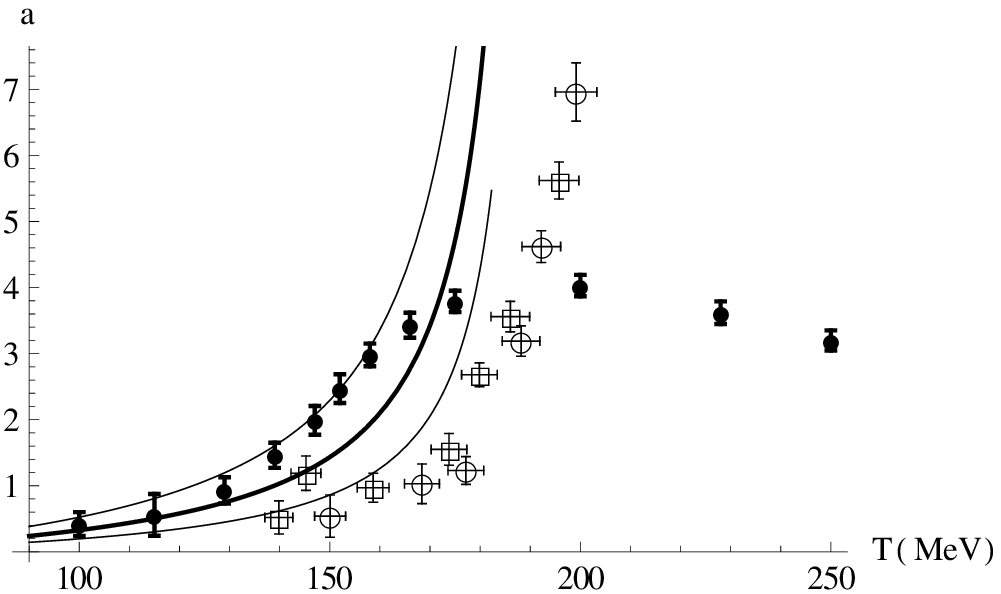}}
                \subfigure[]{\includegraphics[width=0.47\textwidth]{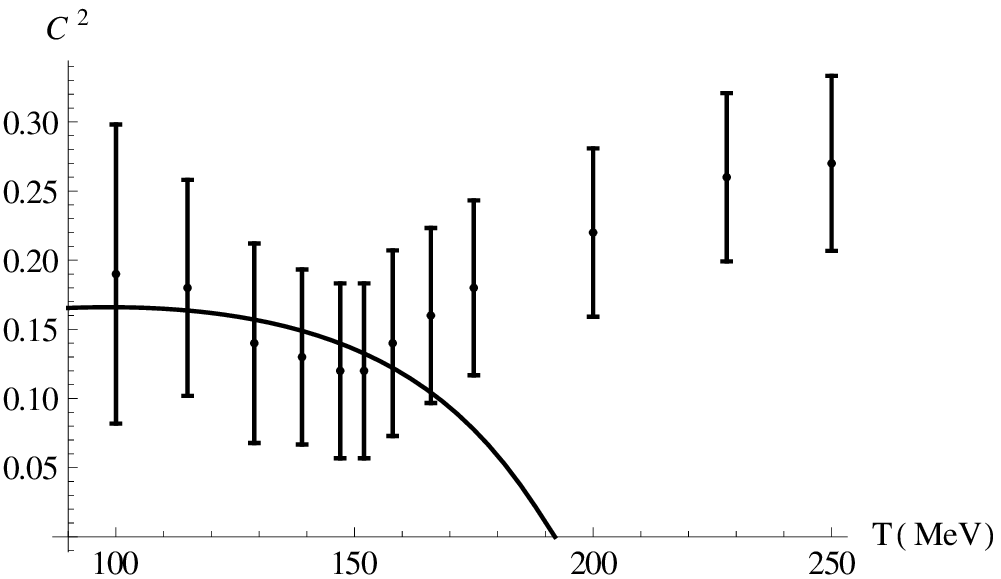}}
               \quad
           \caption{Trace-anomaly (a) and squared sound speed (b) as a function of the temperature. Symbols and lines have the same meaning as in Fig.~\ref{figpa}}
       \label{figcvc2}
   \end{figure}

In Fig.~\ref{figpa}(b) the normalized energy density is compared to L2 data. In this case there is a larger deviation of lattice results in respect with the theory, although the overall shape of curve and data are quite similar in the region up to $T=$200~MeV. This deviation may be attributed, at least partially, to the use of non-physical quark masses. In fact, it is known that the critical temperature from lattice calculations decreases as the quark masses approach the physical values~\cite{Arriola,FKarsch2003b}. It was shown~\cite{Arriola} that the use of physical quark masses would result in a shift in the critical temperature towards low temperatures of about 20~MeV. It is possible to see in Figs.~\ref{figpa} and~\ref{figcvc2} that such a shift would lead to a good agreement between L2 data and theory.

 In Fig.~\ref{figcvc2}(a) the trace-anomaly is reported. One can see that up to $T\sim$175~MeV the theoretical calculations fall between the two kinds of lattice-QCD results, being systematically above the results from L2 group and systematically below the results from L1. Considering the uncertainties in the parameter $\gamma$, it is possible to find an agreement with the calculations from Ref.~\cite{Zoltan_Fodor}, but it is not possible to discard the agreement with those from Ref.~\cite{F_Karsch}. Despite the same critical temperature, the deviation between thermodynamical and lattice calculations starts at lower temperatures in the case of the trace-anomaly, reflecting the higher sensitivity of the latter to the critical temperature.

The speed of sound in hadronic medium is plotted in Fig.~\ref{figcvc2}(b) and compared with  L1 calculations. Now a nice agreement is found between lattice calculations and theoretical results due not only to the larger uncertainties in lattice data but also to the fact that the speed of sound is independent of $\gamma$, which is the most uncertain parameter in the theory. As it was expected~\cite{Castorina} the speed of sound goes to zero at the critical temperature.

In~\cite{Biro2,Biro3} it was shown that finite-size effects can give rise to non-extensivity and  that the ratio $\tau/T$ should be proportional to $\exp{(-S/C)}$, where $S$ is the entropy and $C$ the heat capacity. Using $C_V$ for $C$ and observing the practically constant speedy of sound in Fig.~\ref{figcvc2}(b), we see that the linear property for $\tau(T)$ used here is in qualitative agreement with the results in~\cite{Biro2,Biro3}.

It is not the objective of this work to determine which of the lattice-QCD calculations are in better agreement with the non-extensive self-consistent theory. For this purpose more precise determination of the parameters and more accurate calculations are required. Such an accuracy may be achieved, for instance, by considering distinct mass spectra for bosons and fermions. However, the results presented here show that there is a better agreement between the theoretical results and L1 or L2  than between L1 and L2.


The results obtained so far indicate that the non-extensive self-consistent theory is able to directly connect experimental information to lattice-QCD data since it describes on one hand the experimental results for $p_T$-distributions and the observed hadronic states, and on the other hand  the thermodynamical functions as calculated in lattice-QCD. In this sense the theory represents an interesting opportunity to understand the phase transition from confined to deconfined regimes.

It is worth to mention that HRG models can describe quite well the lattice data~\cite{FKarsch2003}. The main difference between the approach presented here and HRG results is that the non-extensive self-consistent theory is directly based on experimental data for $p_T$-distributions, while the latter approach lacks this base.

Concluding, in this work several thermodynamical functions of interest in the study of excited hadronic systems are calculated according to the non-extensive self-consistent thermodynamical theory proposed in Ref.~\cite{Deppman}. A connection between the non-extensive functions and the corresponding lattice-QCD quantities is established. This allows the comparison of the results obtained by using the non-extensive thermodynamics to those obtained in lattice-QCD calculations.

The parameters $T$, $q$  and $\gamma$ used in the thermodynamical theory are those determined directly from experimental $p_T$-distributions or from hadron spectrum~\cite{Lucas}, therefore no free parameters are used to fit the results of lattice QCD. The comparison shows the agreement between theory and lattice calculations for all thermodynamical functions analyzed here.  Also, it shows that a power-law mass spectrum is consistent with lattice results within the non-extensive thermodynamics.

With these results, the non-extensive self-consistent theory shows to be an important tool for the study of thermodynamical properties of hadronic medium. In fact, it completely describes the experimental information from high-energy collisions and from the observed mass spectrum, and on the other hand it is in agreement with the lattice-QCD calculations for the bulk thermodynamic functions.

\section{Ackowledgements}

The author is thankful to E. Meg\'ias, from University of Barcelona, for reading the manuscript and for interesting suggestions.
Also, the author benefited from interesting discussions  with J. Noronha and T. Mendes, from University of S\~ao Paulo, and with  C. Tsallis, and I. Bediaga, from Centro Brasileiro de Pesquisas Físicas. This work received support from the Brazilian agency, CNPq, under grant 305639/2010-2.

\bibliographystyle{aipproc}

\end{document}